# Morphological and Chemical Changes in Cd-free Colloidal QD-LEDs During Operation

Ruiqi Zhang[1,2], Jamie Geng[1,2], Shaun Tan[3], Shreyas Srinivasan[3], Taehyung Kim[4], Mayuran Saravanapavanantham[1,2], Kwang-Hee Lim[4], Mike Dillender[1,2], Heejae Chung[4], Thienan Nguyen[1], Karen Yang[1,2], Yongli Lu[3], Taegon Kim[4], Moungi G. Bawendi[3],

and Vladimir Bulović[1,2,*]

1. Department of Electrical Engineering and Computer Science, Massachusetts Institute of Technology, Cambridge, MA, 02139, US
2. Research Laboratory of Electronics (RLE), Massachusetts Institute of Technology, Cambridge, MA, 02139, US
3. Department of Chemistry, Massachusetts Institute of Technology, Cambridge, MA, 02139, US
4. Samsung Advanced Institute of Technology, Samsung Electronics, Suwon, Republic of Korea
* Corresponding Author

**Abstract**

Heavy metal-free quantum-dot light-emitting devices (QD-LEDs) have demonstrated remarkable brightness, saturated color, and high efficiencies across a broad spectral range. However, in contrast to organic LEDs (OLEDs), QD-LED operational lifetimes remain limited, with the underlying degradation mechanisms not fully understood. In the present study, we show that InP/ZnSe/ZnS (red-emitting) and ZnTeSe/ZnSe/ZnS (blue-emitting) cadmium-free colloidal QD-LEDs undergo nanoscale morphological changes during operation. Specifically, interparticle coarsening and layer thinning are observed in the electron transport layer (ETL) consisting of ZnMgO nanoparticles (NPs), in the QD emissive layer, and in the organic hole transport layer. This is accompanied by the generation and diffusion of compositional oxygen- and hydrogen-radicals throughout the device, with oxygen accumulating at the electrode/ETL interface. Moreover, *in situ* transmission electron microscopy reveals the electron beam exposure, in the presence of hydrogen radicals, accelerates ZnMgO NPs coarsening. To mitigate these degradation pathway, we show that acrylate-based resin-encapsulation treatment stabilize the ETL/QD layers by suppressing the radical formation and halting morphology changes. This approach achieves dramatic stability enhancements, exhibits an 8-fold and 5000-fold lifetime improvement on InP/ZnSe/ZnS and ZnTeSe/ZnSe/ZnS QD-LEDs, respectively. Our findings establish the causal relationships between the morphological degradation, interlayer radical dynamics, and state-of-the-art QD-LEDs instability, providing new insights into a scalable encapsulation treatment that enables efficient and long-lived Cd-free QD-LEDs.

**Main**

**Introduction**

Colloidal quantum dot light emitting diodes (QD-LEDs) are an emerging technology with potential to be used in the next generation of high-brightness displays with saturated color, and wide-color gamut[1-4]. Within last decades, QDs have been used in display technologies as saturated-color optical downconverters in liquid crystal displays and in organic LED (OLED) displays. Developments of electrically excited QDs in QD-LEDs have demonstrated a light-emitting technology that is compatible with flexible substrates[5], that is environmentally benign[6], has fast response times[7], and has efficient light emission from a simpler device architectures that could lead to lower display manufacturing costs[8]. State-of-the-art red and green QD-LEDs have demonstrated the high external quantum efficiency (EQE), approaching the light out-coupling theoretical limit[9-12]. EQE of blue QD-LEDs has also been improved with recent approaches, such as suppressing organic-inorganic interfacial charge leakage[13], and surface engineering[14]. However, for QD-LEDs to be commercially viable, improvements must be made to their operating lifetimes, while maintaining high efficiency, especially for heavy-metal-free QD-LEDs.

QD photoluminescence (PL) efficiency, and balanced charge injection of electrons and holes into the QD layer, are the dominating factors influencing QD-LED EQE. It is found that continuously applying a forward bias to the Cd-free QD-LEDs has multiple drawbacks, including the presence of quantum-confined Stark effect that quenches the device photoluminescence, deteriorated EQE from the excessively charged QDs which enhances Auger recombination[15], charge accumulation in transport layers, non-radiative recombination center formations[2,16,17], retarred carrier-induced electron injection barriers against emission layer (EML) [9,18,19], and the junction space-charge accumulation[19-22]. Thus, to facilitate charge injection and carrier balance, metal (e.g. Mg, Al) doped ZnO nanoparticles (NPs) are commonly selected as electron transport layer (ETL) material, modulating the interlayer band mismatch by raising the conduction band minimum (CBM). This has been shown to increase electron injections rate, suppresses excessive electron flux, and diminishes exciton quenching[23-25]. So far, over 30% QD-LED EQE has been achieved,[26] with high-efficiency QD-LEDs benefiting from the shelf-aging phenomenon, i.e. positive aging while stored after fabrication[27,28]. It has been shown that QD-LEDs containing the InP/ZnSe/ZnS red QD emissive layer, with ZnMgO ETL, exhibit aging over weeks with an initially increasing EQE (referred to as the positive aging) followed by an extended decrease in EQE (negative aging). During this time, ZnMgO nanoparticles (NPs) morphologicaly ripen[29]. With latest approaches introducing water and resin encapsulation treatments, ZnMgO NPs are found to be morphologically stable and retain their n-doping, while further improving the QD-LED performance[27,30]. However, the QD emission layer still suffers from the electric field inhomogeneity, causing potential device degradation[31].

As the operational EQE degradation deteriorates an initially efficient QD-LED, this suggests that under applied electric field, QD-LEDs undergo chemical or morphological modifications that affect QD PL and the injected charge balance. In the present study, we systematically investigate the potential device degradation mechanisms under forward-bias. We investigate the nano-scale morphological changes in the cross-sectional structures of red and blue Cd-free QD-LEDs comprised of InP/ZnSe/ZnS and ZnTeSe/ZnSe/ZnS QDs, respectively. We observe that progressive device aging induces densification across all functional layers of a QD-LED, resulting in layer thinning and structural rearrangements that correlate with device efficiency loss. By tracing the atomic composition of each QD-LED layer, we identify elemental motility under device bias. To assess the changes in the chemical composition of QD-LED layers as they undergo morphological changes, we perform *in situ* environmental transition electron microscopy (TEM) on drop-casted ZnMgO NPs thin films, and on blue QD-LEDs lamella cross-sections under hydrogen doping. We show that the prescence of hydrogen radicals accelerates coarsening of ZnMgO NPs both in the neat films, and within the ETL of blue QD-LED cross-sections. In contrast, when acrylate-based resin is used as an encapsulation treatment, both red and blue QD-LEDs maintain a stabilized ETL/QD emission layer morphology, and it appears that the radical formation is suppressed, diminishing the interpartical coarsening and layer rippening. This treatment exhibits a large operational lifetime improvement on both red and blue QD-LEDs.

## Results:

## Cd-free QD-LEDs Synthesis and Electro-optical Properties

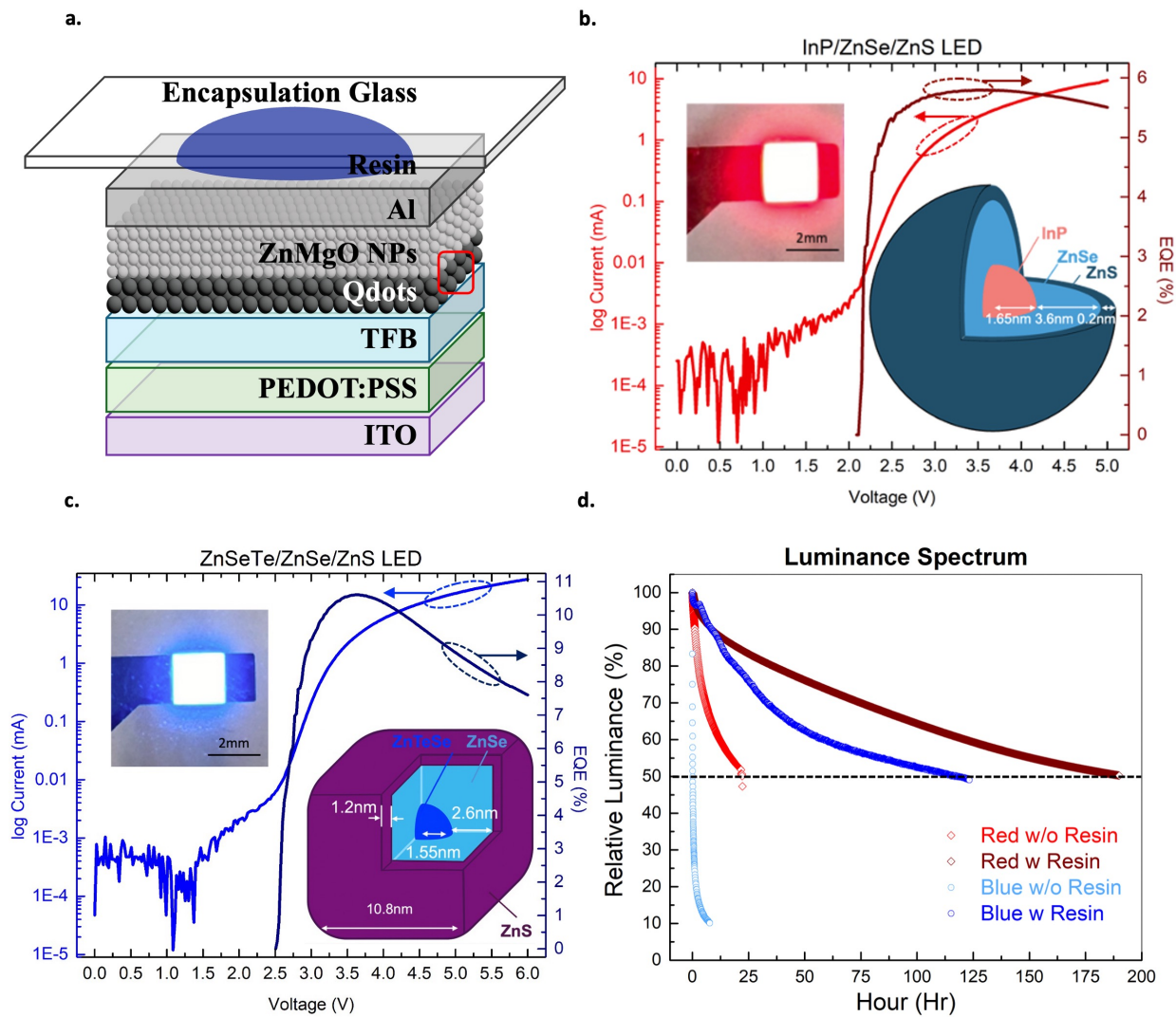

**Figure 1. QD-LEDs Structure and Properties.** | **a**. Cd-free QD-LEDs basic structures, consisting of ITO/ PEDOT:PSS/TFB/QDs/ZnMgO NPs/resin/Al. **b-c.** Operational electronic properties of **b.** InP/ZnSe/ZnS QD-LEDs, and **c.** TeZnSe/ZnSe/ZnS QD-LEDs. Current-voltage spectrum of QD-LEDs under forward biasing is shown on the left y-axis of the spectrum where current is plotted in semilog scale (light red and light blue). Right y-axis indicates the Lambertian calculated voltage- % quantum efficiency (QE) spectrum (dark red and dark blue). Inserted images indicates the functional device with their core/shell/shell (C/S/S) structure and radius

length. Red QD core/shell/shell structures consist of InP, ZnSe, ZnS with radius of 1.65 nm, 3.6 nm, and 0.2 nm, respectively. Blue QD core/shell/shell structures in consist of ZnTeSe, ZnSe, ZnS with radius of 1.55 nm, 2.6 nm, and 1.2 nm, respectively. **d.** Acrylate-based resin influence on LED operation lifetime. Prominent lifetime enhancement is achieved.

Figure 1a and supplementary Figure 1 (Fig.S1) illustrate the typical Cd-free QD-LED consisting of ITO/PEDOT:PSS (hole injection layer, HIL)/TFB (hole transport layer, HTL)/QDs emission layer (EML)/ZnMgO NPs (ETL)/Al, with emission layer composed of InP/ZnSe/ZnS QDs (simplified as InP) and ZnTeSe/ZnSe/ZnS QDs (simplified as ZnTeSe), respectively. Followed by the core-shell consecutive precursor nucleation and purification procedure[3,9], InP QD core (r = 1.65 nm) is formed to be uniformly spherical and covered with the tailored ZnSe (r = 3.6 nm) and ZnS (r = 0.2 nm) shells, suppressing the non-radiative Auger combination. ZnTeSe-based QDs, tend to form quasi-cubic/diamond shapes, with a ZnTeSe core, ZnSe inner shell and ZnS outer shell with radius thickness of 1.55 nm, 2.6 nm, and 1.2 nm, respectively (as sketched in Figure 1b and 1c). Emission wavelength of the Te-doped ZnSe QD core in blue QD-LEDs is tuned by adjusting the Te/Se composition ratio, with an addition of hydrofluoric acid (HF) exposure to suppress the stacking faults during QD shell growing process[9]. Application of forward bias leads to QD-LED electroluminescence (EL), with InP QD-LEDs turning on at 2.1 V, with a peak EL emission wavelength at $\lambda$ = 630 nm (corresponding to the peak energy of 1.97 eV, with FWHM of 110 meV). ZnTeSe QD-LEDs turn on at 2.5 V, with a peak EL emission wavelength at $\lambda$ = 465 nm (2.67 eV peak energy, FWHM of 250 meV). The EQE after turn-on of resin-free red and blue QD-LEDs is measured as 5.8% and 10.6%, respectively, obtained with current-voltage-luminance (JVL) measurement that assumes Lambertian QD-LED emission[32].

To investigate the aging effect, the fabricated red and blue QD-LEDs are subjected to an applied bias, adjusted to maintain a constant device current. The current is then maintained until the device brightness decrease to 50% of its initial value, which happens after LT50 hours of operation (LT50 refers to the aging time needed to reach 50% of the initial brightness). The as-fabricated InP QD-LEDs has an LT50 of 22.1 hours (with initial condition set as 3.29 V, 2.0 mA, that generates 6500 Cd/m$^2$ device brightness). The as-fabricated ZnTeSe QD-LED has an LT50 of 0.2 hours (3.35 V, 0.1 mA, 146 Cd/m$^2$). To enhance the device performance, acrylate-based resin post-fabrication encapsulation treatment is applied, leading to an extended T50 lifetime in both InP and ZnTeSe QD-LEDs. Remarkably, resin-encapsulated InP QD-LED has an LT50 of 189.9 hours (3.06 V, 2.0 mA, 6500 Cd/m$^2$) where the ZnTeSe QD-LED LT50 rises to 115.5 hours (3.56V, 0.1 mA, 500 Cd/m$^2$). With an accumulation factor of $1.8^9$, the InP QD-LED has an 8-fold increase the T50 lifetime under 100 Cd/m$^2$, where ZnTeSe QD-LED has an over 5000-fold LT50 increase.

**Device Nanomorphology Variations and Interlayer Elemental Tracing**

To understand the spatially-resolved degradation signatures and the merits of resin encapsulation, pristine red & blue QD-LED, T50-aged red QD-LED and T70-aged blue QD-LED cross-sectional lamellas are fabricated as sister devices, and prepared with focused ion beam (FIB), as shown in Fig.S2. Lamella thickness is controlled to be less than 200 nm to maximize signal-to-noise level. To accurately analyze interlayer thickness variations, we develop a quantitative algorithm, shown in Figure 2a. A raw TEM image is extracted into column vectors with each vector representing their column-wise grayscale level, ranging from 0 to 255 (left red line). It is then horizontally summed to extract the pixel-wise thickness curves based on the image dimension (right bulk red lines). Each curve is calculated with its neighboring pointwise slopes (derivatives) to find their center index where the gray level encounters the abrupt change, representing the edges of different layers, shown in Figure 2a and Fig.S3. The final multilayer thickness profiles (blue curve with white-crosses, with ±0.5 nm variation) are averaged to get the overall layer thickness, calibrated against the reference scale bar, shown in Figure 2b.

Figure 2b shows a coarsening and compression in the QD layer, together with the ZnMgO NPs layer on the LT-aged InP device. Meanwhile, neighboring QDs exhibit blurred boundaries compared with their pristine counterparts. Quantitatively, a 4.8% post-aging thickness reduction in ZnMgO layer and a 19.5% decrease in QDs layer are observed. Notably, a clear bi-layer QD layer is observed through the gray-level thickness contrast analysis in pristine QD-LEDs. However, this contrast disappears after applying bias to the LED, showing that the QD layer has been coarsened, with QDs merged with each other, losing the distinct boundaries of QDs as seen in the undriven device, Figure 2c. These bias-induced morphological changes affect the QD's core/shell structure, and we would expect they influence the ligand-binding conditions that help to account for bias-induced potential energy shift[33], QDs deterioration[10,21,34] and increased exciton quenching[19] in a degraded devices.

To probe the bias-induced interlayer elemental diffusions, spacially-resolved composition mapping is collected using energy dispersive X-ray spectroscopy (EDS). Figure 2d and Fig.S4 show the essential oxygen EDS mapping profile across the InP QD-LED cross-section (also see Fig.S5). A significant increase in oxygen peak intensity at the ZnMgO/Al interface appears on the aged device, suggesting a formation of an oxidized Al layer after device storage and biasing, which would lead to an increase in device series resistance. Meanwhile, an extra oxygen peak is also present in the aged QD layer, suggesting that the oxygen migrates across the neighbouring layers into the QD layer. The extra oxygen peak could, for example, contribute to the operation-induced charge balance improvement on aged QD-LEDs due to the degradation of electron-injection capability at the ETL/cathod interface[20].

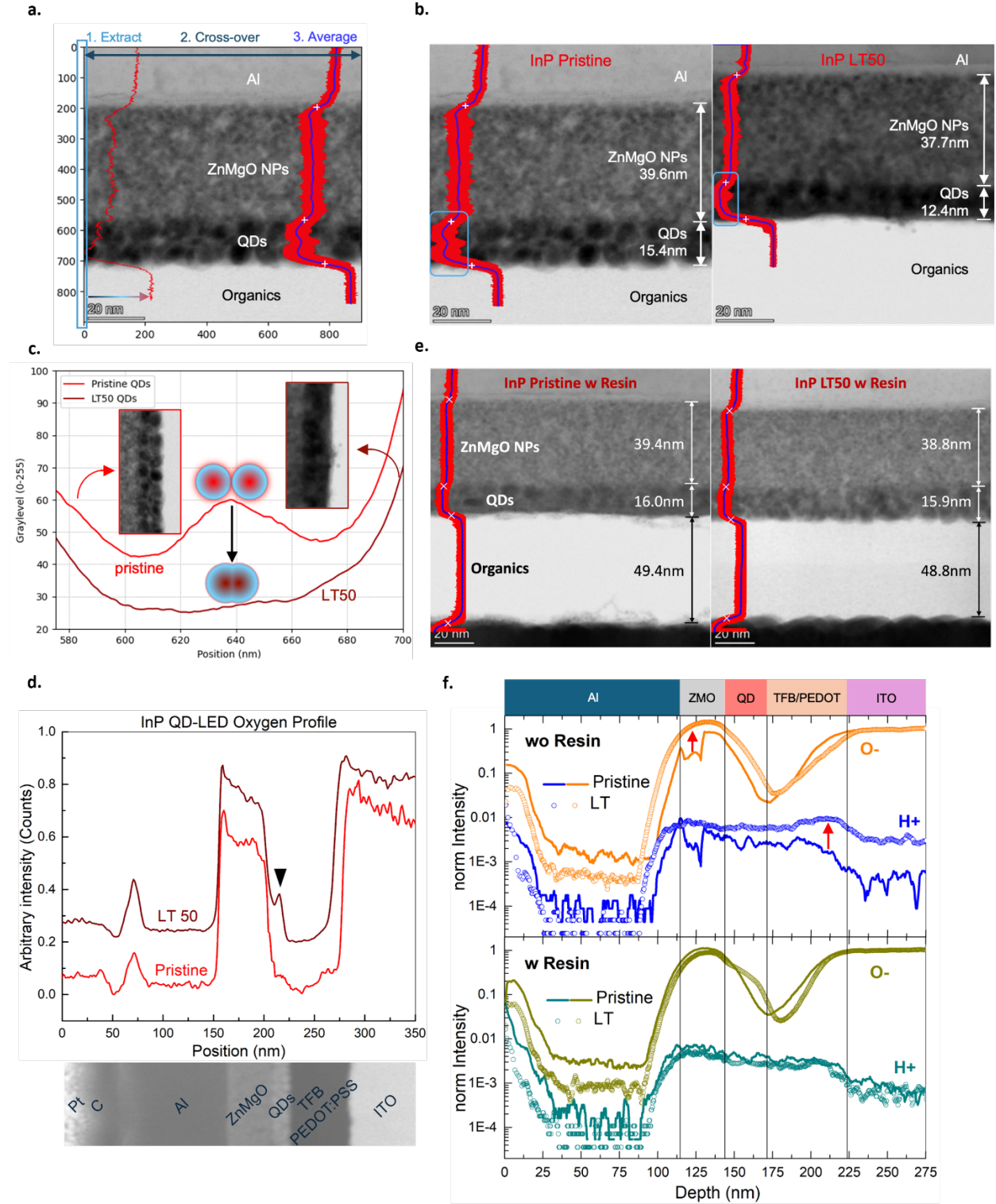

**Figure 2. InP QD-LED Degradation Characterization with the influence of Acrylate-based resin.** | **a**. Computational quantitative analysis process of a QD-LED cross-sectional TEM image. Composition of each layer is labeled in white. Three sequential steps are: 1. Columnwise pixel extraction; 2. Image horizontally cross-over; 3. Find slope index and take average (i.e. vertical-axis index around 200, 580, and 700 in the image, indicated as white cross markers). Light blue rectangle represents how each column is subtracted into grayscale 2D curve (red single line). Bulk red line on the right portion of the image represents the statistical results across the whole image. **b.** Cross-sectional TEM image of pristine (left) and LT-50 aged (right) InP/ZnSe/ZnS QD-LED. **c.** Gray-scale level of the zoomed in InP/ZnSe/ZnS quantum dots layer versus depth position spectrum. The QD gray-scale level are obtained with the region that is squared in light blue rectangle in **b** above. Schematics of two QDs inserted inside the spectrum mimic how two layers of red QDs are coarsened and merged into one condensed layer after operational device degradation. **d.** Energy dispersive X-ray Spectroscopy (EDS) Oxygen spectrum of red InP QD-LED with its cross-sectional TEM image inserted below. Prinstine and LT-aged curves are plotted in light and dark red color, correspondingly. An extra oxygen peak is presented in InP/ZnSe/ZnS QD layer after operational biasing, indicating the operational-induced oxygen migration. At the Al/ZnMgO junction, the spectrum indicates an increase in oxygen level in both pristine and aged red QD-LEDs. **e.** Cross-sectional TEM image of pristine (left) and LT-50 aged (right) acrylate-resin-induced InP QD-LEDs. **f.** TOF-SIMS spectrum of $O^-$ and $H^+$ radicals in pristine and LT-50 aged InP QD-LEDs. Lower and upper spectrum represents the QD-LED with and without acrylate-resin treatment, respectively. QD-LED layer information is indicated on top.

In comparison, resin-encapsulated InP QD-LED show a stable interlayer morphology after T50 operational aging, shown in Figure 2e. Meanwhile, the extra oxygen peak in the QD layer is inhibated, as shown in Fig.S6. It is reported that acrylic acid resin ($CH_2$=CHCOOH) is able to react with hydroxyl group (-OH) inside ZnMgO that generates water, chemically absorbing and interacting with the ZnMgO surface[27]. This water-treated ETL introduces the oxygen absorption sites, likely hindering the release of oxygen radicals from ZnMgO layer that could further diffuse into the QD layer.

Time-of-flight secondary ion mass spectrometry (TOF-SIMS) is carried out to investigate aging effects on the compositional radicals, as shown in Fig.S7. To investigate the potential oxidation and reduction behaviors after device aging, figure 2f shows a comparison between $O^-$ and $H^+$ radicals level in the devices with and without resin. The normalized oxygen and hydrogen intensity levels largely increase for the LT50-aged device without resin treatment, especially in the ZnMgO and InP QDs layer. However, the oxygen level decreases in the ZnMgO layer after resin treatment due to the formation of oxygen vacancies in the metal oxide. The shift in $O^-$ radicals peaks at the QDs/HTL layers interface towards the organics on both w/ & w/o resin devices provides an insight that the carrier recombination is more likely to happen at the EML/HTL interface where electrons are easier to inject into the QD layer as compared to the holes, causing a relatively easier morphological changes to the EML/HTL interface as compared to the EML/ETL interface. Moreover, the $H^+$ radical maintains a constant level after resin treatment, indicating that it suppress the potential bias-induced device reduction. As metal cations are able to extract protons from bonded H under applied bias during device aging, neutral H has the possibility to be reduced into $H^-$ radical. Thus, a comprehensive interlayer $H^-$ radical concentration is also investigated. Fig.S8 shows the normalized $H^+/H^-$ radical intensity across the pristine and LT50 aged InP QD-LEDs. Similar to $H^+$ and $O^-$ radicals, resin-treated InP QD-LEDs maintain a similar $H^-$ radical level, where resin-free devices exhibits an increase in both H-radicals across the n-i-p diode region. This further validates that the acrylate-based resin induced stabilization effect, which greatly extends the device lifetime.

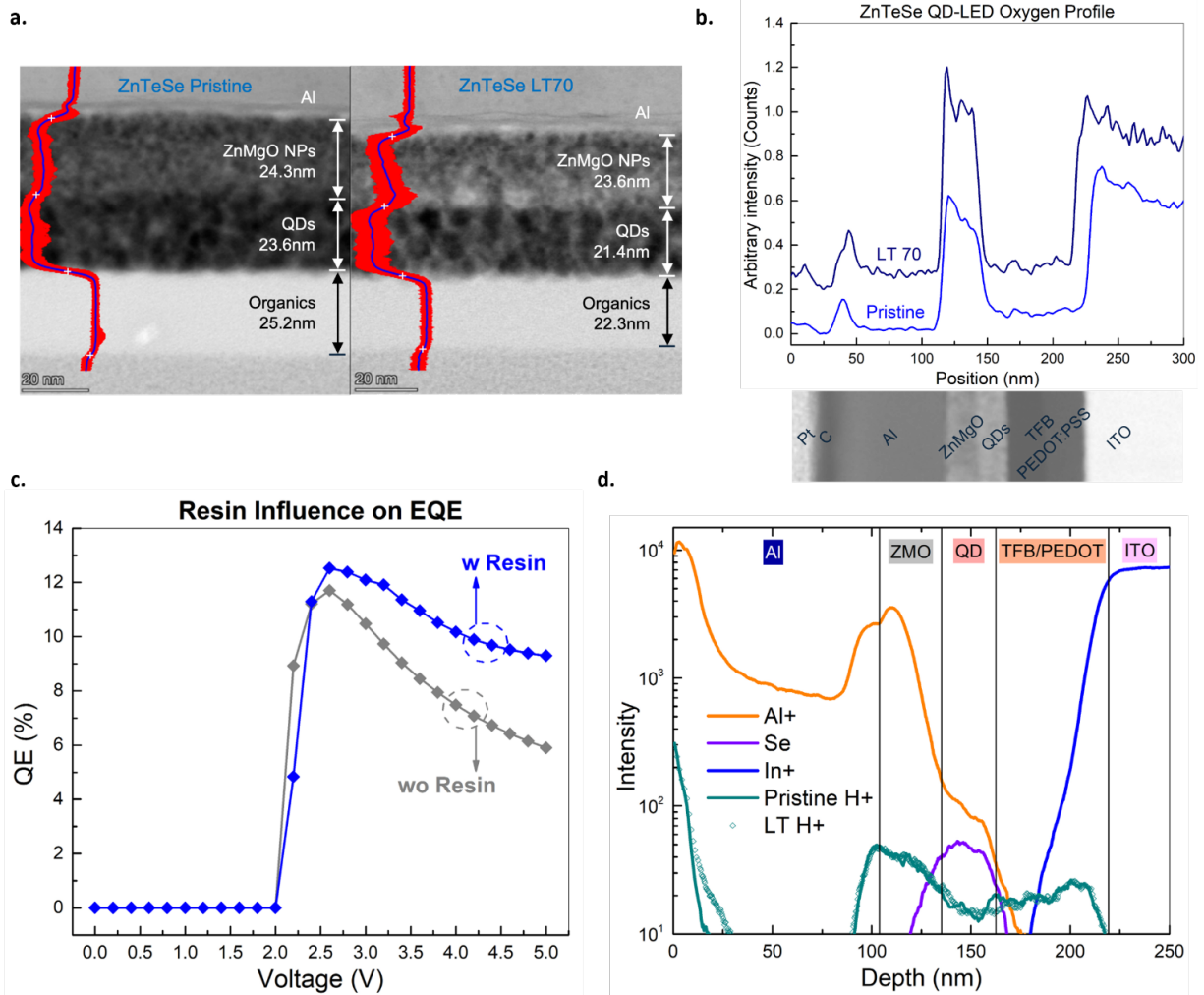

**Figure 3. ZnTeSe QD-LED Degradation Characterization.** | **a**. Cross-sectional TEM image of LT-70 aged resin-free ZnTeSe QD-LED. Thinning layer thickness is observed in all QD, ETL and HTL (organics) layers. **b**. Oxygen EDS spectrum of ZnTeSe QD-LED with their cross-sectional TEM image inserted below. Prinstine and LT-aged curves are plotted in light and dark blue color, correspondingly. The spectrum indicates an increase in oxygen level at the Al/ZnMgO junction. The oxygen level further increases after aging of the device. **c.** Resin influence on ZnTeSe QD-LED EQE. Blue and gray curves represent the device with and without the influence of resin, respectively. **d.** TOF-SIMS spectrum of ZnTeSe QD-LED. Solid line represents pristine device where scatters represent LT50 aged device.

As seen above for the InP QD-LEDs, LT70 aged resin-free ZnTeSe QD-LEDs show a 2.9% reduction in ZnMgO layer thickness and a 9.3% reduction in QDs layer, as indicated in Figure 3a.

A 11.5% thickness reduction is also observed in the organic HTL/HIL layer. In contrast to the InP QD-LED, ZnTeSe QD-LED doesn't show an oxidation of the QD layer, as shown in Figure 3b (also see Fig.S9). Plausible reason includes the rapid decay (0.2h) of resin-free ZnTeSe QD-LED that significantly reduces the total amount of charge that passes through the device, as compared to the much longer lived InP QD-LEDs. Also, of note is the non-spherical self-assebly process of the core/shell/shell structure during blue QD synthesis, which could affect the oxidation of QDs. In comparison, acrylate-resin-encapsulated ZnTeSe QD-LED exhibits an elongation in the LT50 time, similar to the resin-based InP QD-LEDs. Under the same turn-on voltage, Figure 3c illustrates a boost in EQE for the acrylate-resin-encapsulated ZnTeSe QD-LED, consistent with the reported n-doping that increases the conductivity of ZnMgO layer, while further enhancing the device EQE. After T50 degradation, an unchanged TOF-SIMS $H^+$ radical level further validates the effectiveness of resin-treatment in inhibiting elemental diffusions throughout the QD-LED, extending its lifetime, as shown in Figure 3d.

**ZnMgO NPs Ripening and Hydrogen Doping**

It has been shown that aging of the ZnMgO NPs layer causes uncontrollable degradation of Cd-free QD-LEDs[29]. We expect that the exposure of the metal oxide charge transport layer to the chemical radicals could be the cause of the observed morphological modifications under current passage. Thus, we further test the ZnMgO ETL under exposure to the $H^+$ and $O^-$ radicals.

ZnMgO NP films are prepared by drop-casting their colloidal solution onto epoxy-coated TEM grids, enabling a TEM analysis, which we perform under different *in situ* conditions. The electon beam generated from the field emission gun in high-resolution TEM (HR-TEM) induces ionization of residual $O_2$ and $H_2$ gases, so we first assess the effects of TEM beam irradiation on ZnMgO thin film prior to radical exposure.

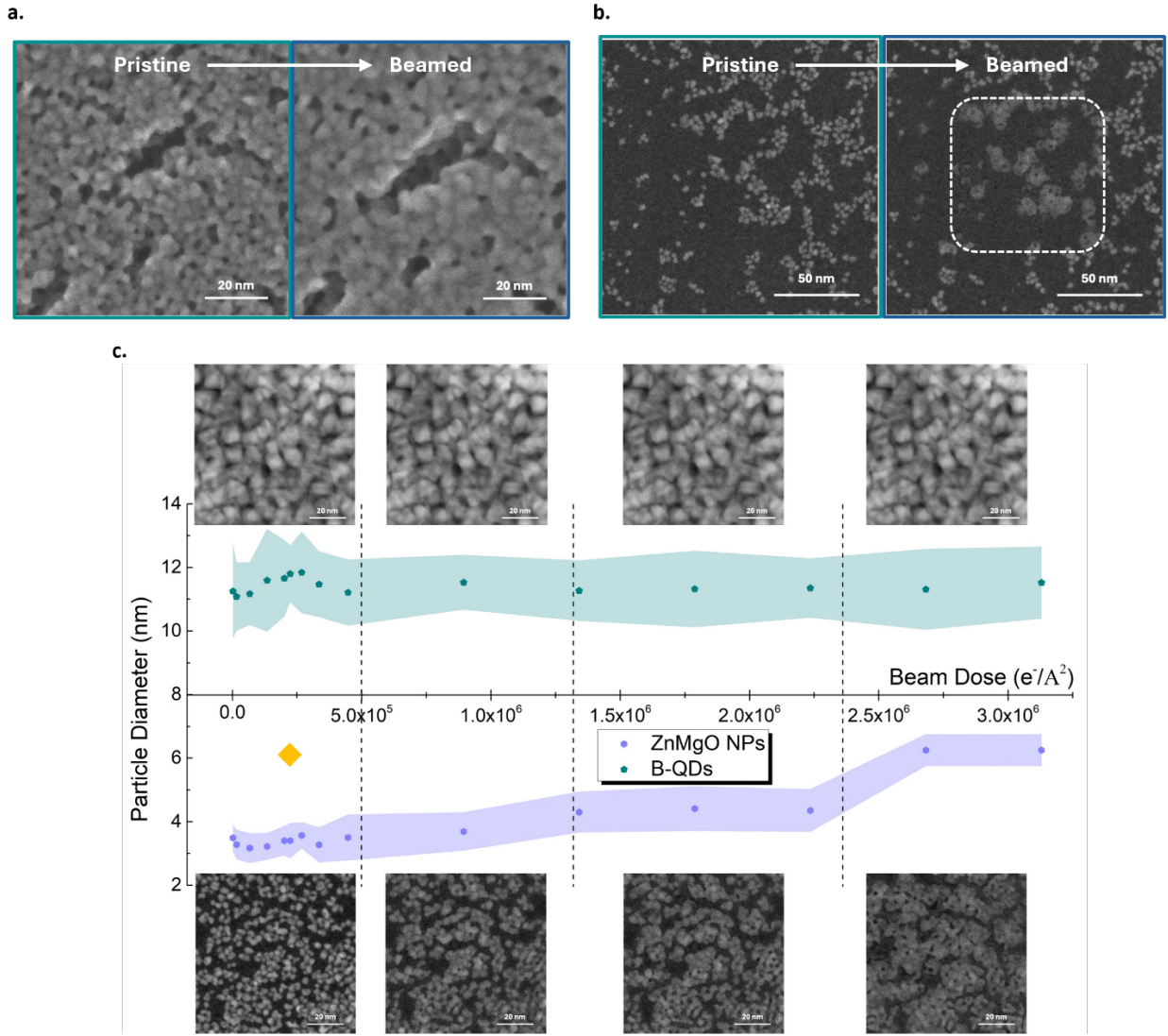

**Figure 4. Influence of Beam Irradiation on ZnMgO NPs and ZnTeSe QDs.** | HR-TEM images of **a**. pristine (left) and beam irradiated (right) ZnMgO NPs thin film acquired in the same region of interests. An increase in nanoparticle size and a coarsening in nanoparticle boundaries are observed. **b**. Influence of beam irradiation on ZnMgO NPs with area of less dense particles. White rectangle indicates the area where electron beam was directed. Similar coarsening effect is observed as in **a** above. **c.** Quantitative analysis of coarsened particle diameter versus total injected beam dose. At each dose, more than 30 randomly chosen particles are measured

to get their averaged diameter. The purple-colored band of measured values represent the scatter plot of particle diameters of ZnMgO NPs. Also shown is the cyan-colored band of measured values representing the scatter plot of particle diameters of ZnTeSe QDs, under various dozes of TEM electron beam irradiation. Solid filling of the area covering the scatter points illustrates the measured error bar at each dose density. Yellow diamond illustrates the dose at which we observed the ZnMgO NP coarsening when hydrogen is presents, as discussed in Figure 5 below.

Pristine ZnMgO thin film is first imaged with acquisition time of 17.6s, limiting the potential electron beam damage. Electron beam is then directed to continually irradiate an area of a film for an extended period of time, which is intended to mimic charge injection and charge transport through the tested film, as would be experienced during LED operation. Figure 4a and Fig.S10 illustrate the influence of beam irradiation on the ZnMgO NPs where after 6 minutes of beam exposure, ZnMgO NPs get coarsened into larger ones, with less defined particle boundaries. The same experiment is repeated on a broader, but less densely packed, ZnMgO NP film. As shown in Figure 4b, circled area has undergone a continuous beam irradiation and exhibits coarsening, with neighbouring NPs merging and evolving from nanoparticle clusters into thin film textures. To quantitatively investigate the relationship between the total electron dose passing through the ZnMgO layer versus the level of coarsening, we calculate the electron beam dose density per unit area based on the TEM's inherent current density, probe size, dose rate, acquisition window frame rate, and irradiation time[38-40], as shown in Figure 4c. We find that as total dose increases from $2.39 \times 10^3 \frac{e^-}{Å^2}$ to $4.47 \times 10^5 \frac{e^-}{Å^2}$, no significant coarsening is observed. However, starting at a higher electron dose of $8.94 \times 10^5 \frac{e^-}{Å^2}$, ZnMgO NPs begin to aggregate and exhibit ripened coarsening. An obvious loss of nanoparticle features appears at a dose higher than $2.24 \times 10^6 \frac{e^-}{Å^2}$ (see Fig. S11,12), demonstrating the degradation impact on ZnMgO NPs due to electron beam current passing through. In a separate experiment, a similar range of TEM doses are applied to a drop-casted ZnTeSe QDs thin film. No obvious morphological changes are observed across the same beam dose range as used for testing of the ZnMgO NPs films.

As previously illustrated (Figure 2f and 3d), both hydrogen and oxygen radicals are detected in InP and ZnTeSe QD-LEDs, regardless of resin encapsulation, indicating radical generation within the ZnMgO ETL and in the adjacent QD emissive layer. Especially, devices without resin treatment exhibit a pronounced increase in radical concentration in the charge injection/recombination region. To assess the influence of hydrogen doping, the drop-casted ZnMgO NPs thin film are *in situ* exposed to $H_2$ gas during TEM imaging. $H_2$ gas is introduced into the TEM chamber at 3 sccm for 6 minutes, exposuring the ZnMgO-coated TEM grid to hydrogen-rich conditions. The TEM beam is coherently turned on, generating the H-radicals. We observe that under the exposure to both H-radicals and the electron beam, ZnMgO NPs exhibit rapid coarsening, shown in Figure 5a. Notably, with the presense of H-radical, ZnMgO NP thin films start observably coarsening at a lower electron dose at $2.01 \times 10^5 \frac{e^-}{\text{Å}^2}$, a dose level at which no coarsening occurred when no H-radicals were present, indicated as the yellow diamond in Figure 4c. This electron dose corresponds to the cumulative electron exposure experienced by a 2-by-2 mm$^2$ QD-LED after 3.58 hours of operation under a constant driving current of 1mA. Without hydrogen, time to achieve the electron dose that observably coarsens the ZnMgO NPs is 15.9 hours, still substantially shorter than the device operational lifetime. This observation again confirms that the electron beam irradiation leads to changes in the morphology of ZnMgO NPs (that form an ETL layer in a QD-LED with H-radical accelerating this process. Additionally, it appears that in an acrylate-resin encapsulated QD-LEDs, the resin chemically reacts with ZnMgO ETL to inhibit the formation of $H^+$ radicals, minimizing the morphological change of ZnMgO NPs (Figure 2f), which maintains the n-type conductivity and minimizes morphological changes and the resulting carrier leakage[36,37], all leading to a longer operational lifetime for both InP and ZnTeSe QD-LEDs.

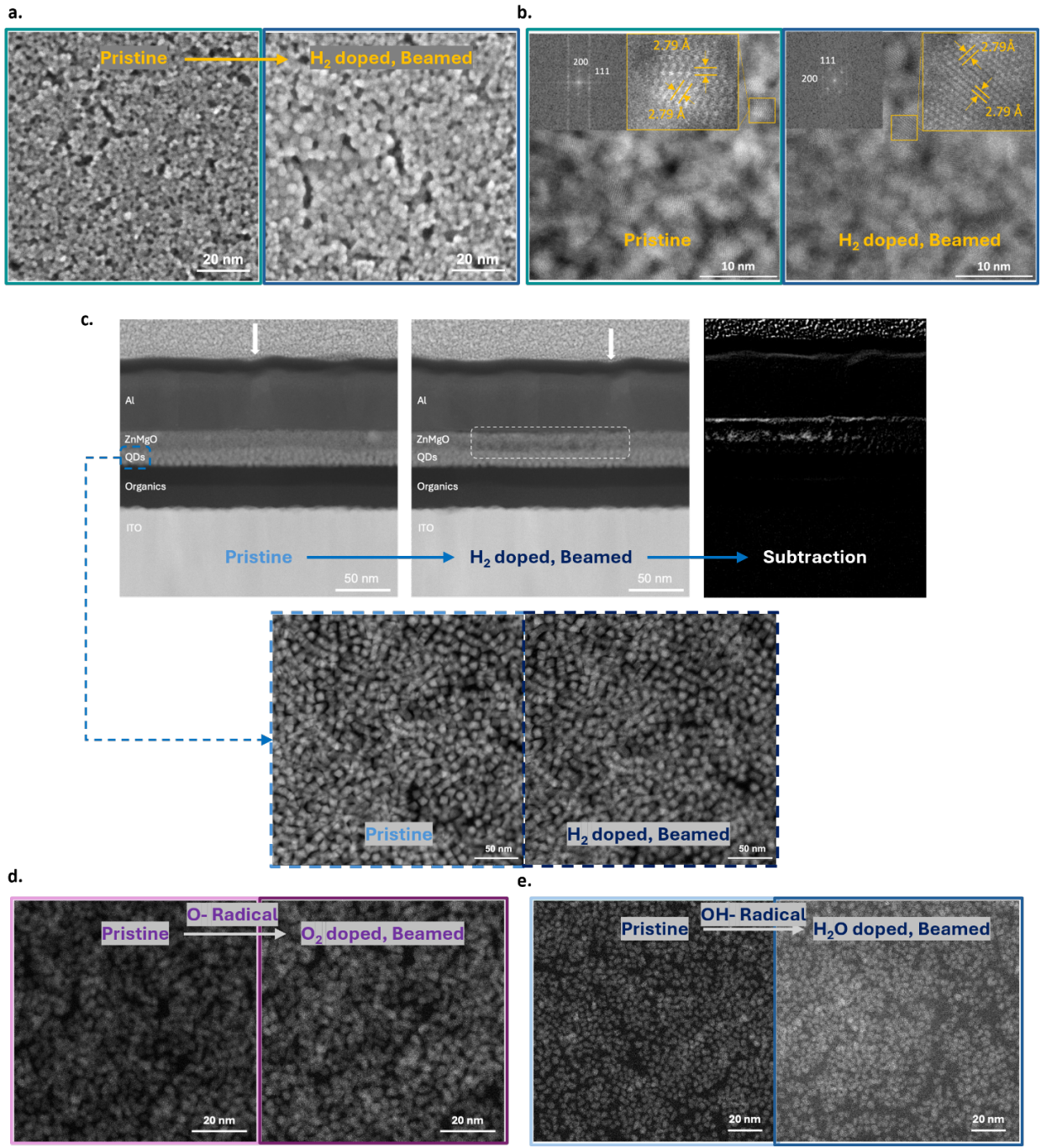

**Figure 5. Influence of Hydrogen Doping on ZnTeSe/ZnSe/ZnS QD-LEDs as well as Oxygen-doping and Water-doping.** | HR-TEM image of **a.** Prinstine (left) and TEM-beam-irradiated & hydrogen-doped

ZnMgO NPs thin film (right). Same region of interest is shown. Similar coarsening and enlarging of NPs are observed with a smaller electron dose density, as compared to the experiments with no H-radicals present. **b.** Atomic resolution HR-TEM images of pristine and hydrogen-doped ZnMgO NPs thin film. Inserted images (upper right) indicate the zoomed-in atomic structure of the selected region. Atomic lattice spacings are measured in both HR-TEM images. Inserted images (upper left) represent the 2D Fast Fourier Transform (FFT) patterns of the atomic structure extracted from the raw image. ZnO (111) and (200) planes are acquired in the in-plane direction. **c.** TEM e-beamed & H-doped ZnTeSe QD-LED cross-sectional lamella. A coarsened ZnMgO NPs layer is observed after TEM electron beam irradiation and hydrogen doping. The white box indicates the area where extended electron beam irradiation was performed after hydrogen doping, showing the NP coarsening. White arrows indicate the same device feature on the lamella, which allows us to subtract the the image before the extended TEM irradiation from the one after the irradiation. HR-TEM images of **d.** beam-irradiated & $O_2$-doped and **e.** beam-irradiated & $H_2O$-doped pristine (left) and doped ZnMgO NPs thin film (right). The same amount of electron dose is applied to ZnMgO NP films doped with $O_2$ and with $H_2O$, as was applied to ZnMgO NP films doped with $H_2$, but no significant coarsening of NPs is observed.

Two-dimensional Fast Fourier transforms (2D-FFT) image analysis is carried out to determine the lattice spacing of ZnMgO NPs. As shown in Figure 5b, atomic-resolution TEM images are taken before and after the electron beam exposure and hydrogen doping. Spatial measurement reveals a 0.28 nm d-spacing in both the pristine and the beam-exposed device, matching the ZnO lattice parameter[41-43]. The corresponding 2D-FFT diffraction patterns exhibits a hexagon symmetry, with four parallel vertices representing the ZnO (111) plane and two corner vertices representing the (200) plane[44-46]. These observations confirm that ZnMgO NPs coarsening does not alter the atomic lattice spacing, nor does it induce in-plane lattice extension or suppression. Instead, the enlarged particles arise from the electron-induced particle growth and the aggregation effect of neighboring ZnMgO NPs, validating the shrink of ETL along out-of-plane direction in LED.

Potential H-doping and beam irradiation effect is further tested on ZnTeSe QD-LED cross-sectional lamella to examine the in-device radical influence. Illustrated in Figure 5c, a 10-minute exposure to $H_2$ (3 sccm) under electron irradiation induces signifcant coarsening of the ZnMgO ETL, whereas the QDs EML and organic HTL remain unaffected. A further pixelwised image subtraction highlights the a significant discrepancy only at the ETL/contact and ETL/EML juctions. To substantiate the post-doping ZnTeSe QDs layer structual consistensy, we applied an identical TEM electron dose desnity and $H_2$-concentration doping treatment to a drop-casted ZnTeSe QDs thin film on a TEM grid. Figure 5c depicts the pristine and beam/$H_2$-treated blue-QDs film, respectively, where no changes are observed in the QD film. (see Fig.S13-15 for additional proof). Further computational particle-size analysis is performed by applying a Gaussian filter that enhances edge detection to distinguish the ZnTeSe particle size, shown in Fig.S16. The average QD radii in pristine (5.90±0.5 nm) and doped (5.88±0.5 nm) films are indistinguishable within the error. This is consistent with the overall measured mean radii (5.98 $\pm$ 0.5 nm), confirming the resilience of QD films to electron doses and $H_2$-doping. Hence, ZnMgO NPs are shown to exhibit aggregation and coarsening under exposure to current flow, filling pre-existing voids and reducing the effective ETL thickness as compared to the pristine device.

To test the influence of $O_2$-doping and $H_2O$-doping on the morphology of the ZnMgO NP layers, we tested the morphology of ZnMgO NP films before and after the electron-beam-irradiation that followed the $O_2$-doping and $H_2O$-doping conditions, with data shown in Figure 2f. After applying the same TEM electron dose amount and an identical gas doping, in contrast to H-radical which accelerates the morphological change in the ZnMgO NPs, both O- and OH-radicals present no obvious changes to the ZnMgO NP layer, as shown in Figures 5d, and 5e. Thus, we have shown the dominating role of the electron beam irradiation and H-radical in Cd-free QD-LEDs morhological change, which might be responsible for the QD-LED degradation.

**Conclusion**

In summary, as progress toward the commercial viability of heavy-metal-free QD-LEDs advances, our work provides a thorough understanding of device degradation mechanisms under operational bias. Through systematic analysis of evolution of device operations, we, for the first time, identify the shrinking and coarsening behaviors within InP and ZnTeSe Cd-free QDs emission layers, metal oxide ETL, and organic HTL in a QD-LED under LT-aging. The detailed interlayer compositional analysis clearly indicates the consequential impact on radical-induced device chemical circumstances variations to the decrease of device efficiency and operation lifetime, through disturbing the charge injection and recombination rate and particle stabilities. In contrast to O- and OH-radicals, exposing ZnMgO NPs to H-radical and electron beam irradiation causes NPs to morphologically coarsen, contributing to the layer thinning and coarsening in aged QD-LEDs. To improve the device operation lifetime, we further demonstrate the acrylate-based resin-encapsulation treatment successfully interacts with ZnMgO NPs ETL, inhibiting the morphologic changes. An 8-fold and 5000-fold lifetime improvement on InP/ZnSe/ZnS red and ZnTeSe/ZnSe/ZnS blue QD-LEDs is observed, respectively.

Overall, we have established a thorough framework in understanding the degradation mechanisms of both state-of-the-art InP/ZnSe/ZnS and ZnTeSe/ZnSe/ZnS QD-LEDs, which we expect is crucial in understanding and improving their long-term stability and performance. By providing the resin-treatment solutions, we have further developed the acrylate-based resin-induced encapsulation in largely enhancing device performance. These findings clarify operational-induced changes in device architecture, offering valuable guidance for diagnosing Cd-free QD-LED degradation and informing future QD-LED designs and optimizations.

## Methods

**Materials and QD-LED Fabrications**

All materials used in this study were purchased for direct use without further purification. Poly(3,4-ethylenedioxythiophene):poly(styrenesulfonate) (PEDOT:PSS (AI 4083)) was dissolved in aqueous solution and purchased from Ossila. Materials were kept refrigerated during storage. Poly(9,9-dioctylfluorene-alt-N-(4-sec-butylphenyl)-diphenylamine) (TFB) was purchased from Montreal Optoelectronics Inc. with purity of >99%. O-Xylene (anhydrous, 97%, 1,2-Dimethylbenzene) and Oleic Acid (technical grade, 90%) were purchased from Sigma-Aldrich. ZnMgO nanoparticles in ethanol for usage as InP QD-LED electron transport layers and acrylate-based resin for encapsulation treatment were provided by Samsung Avanced Institute of Technology. Synthesis of ZnMgO NPs solution followed by their published literature[9]. These nanoparticles were stored in nitrogen glovebox freezer to prevent aggregation. Aluminum pellets used in the deposition of electrode materials were purchased from Kurt J. Lesker. InP/ZnSe/ZnS QD-LEDs and ZnTeSe/ZnSe/ZnS QD-LEDs for device characterizations were fabricated and encapsulated at Samsung Advanced Institute of Technology (SAIT), following the as cited methods [3,9].

**QD-LED Electro-optical Characterizations**

QD-LED current-voltage (JV) measurement was carried out using a Keithley 2636A source meter. Luminance-voltage (LV) measurement was measured through FDS1010 Si photodiode purchased from Thorlabs Inc. The photodiode was calibrated with its responsivity before use. The photodiode has a rise time of 65 ns, a wavelength of 350-1100 nm with a 10 mm × 10 mm active area. Same Keithley 2636A was used in providing the voltage during LV measurement.

For photoluminescence measurements, a λ=405 nm pulsed laser (PicoQuant, LDH-P-C-405, 2 MHz, with laser pulse duration < 100 ps) or a λ=532 nm pulsed laser (PicoQuant, P-FA-530XL, 2 MHz, with laser pulse duration < 100 ps) was focused with a lens (Thorlabs, LA1978-A-ML) onto the back-focal plane of a 50x 0.7 NA air objective in an inverted microscope (Nikon, Ti-U) to generate a psuedo-wide-field excitation spot of 50 µm in diameter at the HTL/QD interface of the QD-LED. The emitted light was filtered either through a λ=405 nm long pass dichroic filter (Semrock, Di03-R405) or a λ=532 nm long pass dichroic filter (Semrock, Di02-R532) before being coupled in free-space into a spectrometer with a 150 g/mm grating (Princeton Instruments, SpectraPro-300i). For electroluminescence measurements, a constant current density of 10 mA/cm$^2$ was applied to the QD-LED and no spectral filters were used. All spectra were collected with an integration time of 1000 ms.

**Focus Ion Beam and High-resolution Transmission Electron Microscope**

InP/ZnSe/ZnS and ZnTeSe/ZnSe/ZnS QD-LEDs cross-sectional lamellas are prepared with the Raith VELION Focus Ion Beam Scanning Electron Microscope (FIB-SEM) at MIT.nano. Epoxy-encapsulated devices are removed and tailored into quarter-inch size. Tailored device were etched through FIB undergoing the process of encapsulating coverslip removal, 5 nm Carbon-coating, Pt coating, trenching, under-cutting, nano-manipulator inserting, Pt connection, lifting up, TEM grid driving, attaching, Pt coating, lamella milling, lamella thinning and polishing, (again see Fig.S2). Lamellas were etched using a gold ion beam at 35 kV for milling and 5 kV for polishing.

The milled LED lamella is then transferred into the Thermo Fisher Scientific (TFS) Themis Z G3 aberration-corrected scanning transmission electron microscope (STEM) at MIT.nano for imaging. The images were captured under a current of 100 pA with a probe size of 0.7 Å.

**Time-of-flight Secondary Ion Mass Spectrometry**

Positive high mass resolution depth profiles were performed using a TOF-SIMS NCS instrument, which combines a TOF.SIMS5 instrument (ION-TOF GmbH, Münster, Germany) and an in-situ Scanning Probe Microscope (NanoScan, Switzerland) at Shared Equipment Authority from Rice University. The analysis field of view was 100 x 100 μm$^2$ (Bi$^{3+}$ @ 30 keV, 0.3 pA) with a raster of 128 by 128 along the depth profile. A charge compensation with an electron flood gun has been applied during the analysis. An adjustment of the charge effects has been operated using a surface potential. The cycle times were fixed to 100 μs (corresponding to m/z = 0 – 911 a.m.u mass range). The sputtering raster was 450 x 450 μm$^2$ (Cs$^+$ @ 1 keV, 85 nA). The beams were operated in non-interlaced mode, alternating 1 analysis cycle and 3 frames of sputtering per cycle followed by a pause of 3s for the charge compensation. The depth calibrations have been established using the interface tool in SurfaceLab version 7.5 software from ION-TOF GmbH to identify the different interfaces and based on the measured thicknesses prior to the analysis.

**In-situ Transmission Electron Mircoscope and Hydrogen Doping**

Sample Preparation: In preparation for *in situ* TEM imaging, the nanoparticles were removed from the freexer and drop-casted on a Ted Pella copper TEM grid with ultrathin carbon film on lacey carbon. The grids were baked in a vacuum oven at 50°C for 10 hours to evaporate the ethanol solvent and to prevent contamination during TEM imaging.

*In situ* Transmission Electron Microscope: Samples were exposed to hydrogen gas and imaged on a Hitachi HF500 environmental TEM. Without hydrogen gas in the system, the vacuum level of the imaging chamber was $1.7 \times 10^{-5}$ Pascals. Hydrogen gas was flowed into the imaging chamber at 1 to 3 SCCM. With hydrogen in the chamber, the chamber pressure increased to around $10^{-2}$ Pascals. The beam current and voltage ware 15 pA / 100 pA and 200 kV, respectively for all measurements. Microscope probe size is 0.8 Å in diameter.

**Total Electron Dose Calculation and Relevance Device Current**

Total electron dose was calculated following the equation: $R \left[\frac{e^-}{Å^2 s}\right] = \frac{J \left[\frac{C}{s}\right]}{e^- \left[\frac{C}{e^-}\right] \times A [Å^2]}$ where R represents the electron beam dose rate, J represents the microscope current density, in the unit of pA or colume per second, and A represents the total probe size. Probe size was calculated as a square shape. The total electron dose per image $\left[\frac{e^-}{Å^2}\right] = R \times \frac{t}{per\ pixel^2}$. Here, t is the total time where electron beam passing through per area pixel. Same equation is applied when linking electron dose causing ZnMgO NPs to coarsen to the total current applied to QD-LEDs during operation. The total amount of charge that flowed through an aged QD-LED follows: $Q \left[\frac{e^-}{Å^2}\right] =$

$$\frac{I\left[\frac{C}{s}\right]}{1e^-\left[\frac{C}{e^-}\right] \times A\,[\text{Å}^2]} \times t_{driven}[s],$$ where I is the constant current applied to the device, A is the QD-LED operation area, and $t_{driven}$ is the total amount of operational aging time.

**Lambertian EQE Calculation**

The Lambertian EQE calculation follows the cited method[32]: $EQE = \frac{P_{out}/(\frac{hc}{\lambda_{peak}})}{J_{LED}/q}$ where $P_{out}[W] = \frac{L[A]}{f \times R\left[\frac{A}{W}\right]}$ and $f$ is the directionality correlation factor. From the J-V-L set-up, the distance between the photo diode and the LED is $d_{PD} = 50\ mm$. The directionality correlation factor is calculated through $f = \frac{P_{PD}}{P_{out}} = \frac{\int_0^{\theta_{PD}} I(\theta)d\theta}{\int_0^{\frac{\pi}{2}} I(\theta)d\theta} = sin^2\left(\arctan\left(\frac{r_{PD}}{d_{PD}}\right)\right)$. The power of photodiode is obtained through R, responsivity, $P_{PD}[W] = \frac{L[A]}{R\left[\frac{A}{W}\right]}$ @ 630 nm for red and @ 475 nm for blue. The LED output power is calculated through $P_{out}[W] = \frac{P_{PD}[W]}{f}$, and the final EQE follows the equation: $EQE = \frac{P_{out}/E_{Avg}}{J_{LED}/q}$ where $E_{Avg} = \frac{hc}{\lambda_{peak}}$.

**Acknowledgements**

We thank Dr. Aubrey Penn, and Dr. Juan Ferrara from MIT.nano in assisting us in acquiring the HR-TEM images and FIB lamellas. We thank Dr. Yang Yu for FIB training and lamella preparation procedures. We thank Dr. Tanguy Terlier from Rice University for ToF-SIMS measurements. We thank Tamar Kadosh, Jeremiah Mwaura, Tong Dang, Stella Lessler, and Zhangqi Zheng, from MIT Organic and Nanostructured Electronics Lab (ONE Lab) for their many discussions. We thank all our colleagues in Moungi Bawendi's lab for valuable discussions. This work was supported by the Samsung Advanced Institute of Technology (SAIT) (Project Title: Fundamental Understanding of QD-LED Operation and Degradation: Thin Film and Device Optimization). We thank Dr. Taeho Kim, and all of our colleagues from SAIT. The microfabrication and characterization in this work was in part performed at MIT.nano and at MIT ONE Lab (MIT Department of Electrical Engineering and Computer Science).

**Author Information**

Authors and Affliliations

-- Department of Electrical Engineering and Computer Science, Massachusetts Institute of Technology, Cambridge, MA, USA

Ruiqi Zhang, Jamie Geng, Mayuran Saravanapavanantham, Mike Dillendar, Thienan Nguyen, Karen Yang, Vladimir Bulović

-- Department of Chemistry, Massachusetts Institute of Technology, Cambridge, MA, USA

Shaun Tan, Shreyas Srinivasan, Yongli Lu, Moungi G Bawendi

-- Research Laboratory of Electronics (RLE), Massachusetts Institute of Technology, Cambridge, MA, USA

Ruiqi Zhang, Jamie Geng, Mayuran Saravanapavanantham, Mike Dillendar, Thienan Nguyen, Karen Yang, Vladimir Bulović

--Samsung Advanced Institute of Technology, Samsung Electronics, Suwon, Republic of Korea

Taehyung Kim, Kwang-Hee Lim, Heejae Chung, Taegon Kim

Author Contributions

V.B. and R.Z. conceived the idea. R.Z., J.G., V.B., and M.B. developed and optimized the idea. T.K., K.L, and R.Z. contributed to the devices, thin film preparation and aging. R.Z., and J.G. carried out the FIB procedure. R.Z., J.G., and M.S. captured the cross-sectional TEM and In-situ TEM images. R.Z. developed the analytical thickness measurement code and proceeded the FFT transform. T. S, carried out the ToF-SIMS measurements. R.Z., S.S., T.N., J.G., M.S., and K.Y. contributed to the probe station and EL system set-up, PL/EL, JVL, and EQE measurement. J.G., and R.Z. developed the nanoparticle identification python code. M.D., K.Y., and Y.L. assisted with the measurement and optical system set-up. All authors contributed to the discussion of data and commenting on the manuscript.

Corresponding author

Correspondence to Vladimir Bulović (bulovic@mit.edu).